\documentclass[reqno]{amsart}
\usepackage{hyperref}

\begin{document}
\title
{A geometric approach to integrability conditions
 for  Riccati equations}

\author[J. F. Cari\~nena, J. d. Lucas, A. Ramos]
{Jos\'e F. Cari\~nena, Javier de Lucas, Arturo Ramos}  

\address{Jos\'e F. Cari\~nena \newline
Departamento de  F\'{\i}sica Te\'orica, Universidad de Zaragoza,
50009 Zaragoza, Spain}
\email{jfc@unizar.es}

\address{Javier de Lucas \newline
Departamento de  F\'{\i}sica Te\'orica, Universidad de Zaragoza,
50009 Zaragoza, Spain}
\email{dlucas@unizar.es}

\address{Arturo Ramos\newline
Departamento de An\'alisis Econ\'omico, Universidad de Zaragoza,
50005 Zaragoza, Spain}
\email{aramos@unizar.es}

\keywords{Lie systems; Riccati equations; reduction methods; \hfill\break\indent
integrability by quadratures}

\begin{abstract}
 Several instances of integrable  Riccati  equations are analyzed
 from the  geometric perspective of the theory of Lie systems.
 This provides us a unifying viewpoint for previous approaches.
\end{abstract}

\maketitle
\numberwithin{equation}{section}

\section{Introduction}

The Riccati equation
\begin{equation}\label{ricceq}
\frac{dx}{dt}=b_0(t)+b_1(t)x+b_2(t)x^2\,,
\end{equation}
which is a simple nonlinear differential equation,  is very
often used in many  fields of mathematics, control theory and
theoretical physics (see for instance \cite{PW,{CMN}}  and
references therein) and its importance in this field has been
increasing since Witten's introduction of supersymmetric Quantum
Mechanics. It is essentially the only differential equation with
one dependent variable admitting a non-linear superposition
principle \cite{PW,{LS}}. In spite of its apparent simplicity, the
general solution of a Riccati equation cannot be expressed by
means of quadratures except in some very particular cases and
several interesting results about this property can be found, for
instance, in the  books by  Kamke \cite{Kamke} and Murphy
\cite{Mu60}. Few years ago Strelchenya
 presented some integrability conditions
 \cite{Stre}  claiming to complete previous solvability criteria. 
However, it was shown in \cite{CarRam} that such criterion reduced to 
a very well known result, the knowledge of a particular solution. 
Other cases of integrable Riccati equations
appearing in Kamke \cite{Kamke} and Murphy \cite{Mu60}
were also analyzed in \cite{CarRam}. There exist other papers dealing with
  integrability conditions of the Riccati equation \cite{Ra61, AS64, Ra62, RU68}
by reduction to related separable Riccati equations and also recent works
as \cite{Ko06,RDM05, Zh98,Zh99}.

In this paper we review the geometric approach to  the Riccati equation
 according to the results of \cite{CarRam}  with the aim of proving that
the integrability conditions of this equation in the above mentioned cases
 can be understood in a very general way from the point of view of Lie systems.

This paper is organized as follows. In Section 2 we analyze some
known facts about the integrability of Riccati equations. The
geometric interpretation of the general Riccati equation as a
time-dependent vector field in the one-point compactification of
the real line is summarized in Section 3. More specifically, we
study  Riccati equations through an equation in the
group $SL(2,\mathbb{R})$. In Section 4 we describe the action of the
group of curves in $SL(2,\mathbb{R})$ on the set of Riccati equations and
how this action can be seen in terms of transformations of the
corresponding equations in $SL(2,\mathbb{R})$ (for a more geometric
treatment see \cite{CarRam05b}). Finally, in Section 5 we analyze
how our geometric point of view allows us to consider the
integrability conditions as a way to perform a transformation
process from the Riccati equations into integrable ones related,
as Lie systems, with an equation in a Lie subgroup with solvable
Lie algebra. In this way we obtain some known results about
integrability of the Riccati equations found in \cite{Ra61, AS64,
Ra62, RU68, Ko06, RDM05} and we give a theoretical treatment of
why the transformations used there actually work.

\section{Integrability of Riccati equations}

A particular case  for which the Riccati equation is integrable by
quadratures is when $b_2(t)=0$. In this case the equation reduces
to an inhomogeneous linear one and two quadratures allow us to
find the general solution. More explicitly, the general solution
is given by
$$
x(t)=\exp\Big(\int_0^tb_1(s)\,ds\Big)  \Big(x_0+\int_0^t b_0(t')
\exp\Big(- \int_0^{t'} b_1(s) \,ds \Big) dt' \Big) \,.
$$

Note that under the change of variable
$w=-1/x$  the  Riccati  equation
(\ref{ricceq}) becomes a new  Riccati  equation
$$
\frac{dw}{dt}=b_0(t)  w^2-b_1(t) w+b_2(t) \,.
$$
In particular, if in the original equation  $b_0(t)=0$ (Bernoulli for
$n=2$), then the mentioned change of variable transforms the given
equation into a linear one.

On the other hand, the change $v=-x$ transforms the differential equation
(\ref{ricceq}) into a new Riccati equation
\begin{equation}
\frac{dv}{dt}=-b_0(t)+b_1(t)v-b_2(t)v^2\label{ricceq2} \,.
\end{equation}

Another very well-known property on integrability of Riccati equation is
that  if one particular solution $x_1$ of (\ref{ricceq}) is known,
then the change of variable
 $x=x_1+z$ leads to a new Riccati  equation for which  $b_0(t)=0$:
 \begin{equation}
\frac {dz}{dt}=(2  b_2(t)  x_1(t)+ b_1(t)) z+ b_2(t) z^2 \,,
\label{Bereq}
\end{equation}
that can be reduced to an inhomogeneous  linear equation with the change
$z=-1/u$. Consequently, when one particular solution is known, the general
solution can be found by means of two quadratures; $x=x_1-1/u$ with
\begin{align*}
u(t)=&\exp\Big( -\int_0^t[b_1(s)+2 b_2(s) x_1(s)]\,ds\Big) \\
    &\times  \Big(u_0+\int_0^t b_2(t')
        \exp\Big( \int_0^{t'} [b_1(s)+2 b_2(s) x_1(s)]\,ds
\Big) dt'\Big) .
\end{align*}

 When not only one but two particular solutions of (\ref{ricceq}) are known,
$x_1$ and $x_2$, the general solution can be found by means of
only  one quadrature. In fact, the change of variable $x=(x_1-z
x_2)/(1-z)$, or in an equivalent way, $z=(x-x_1)/(x-x_2)$,
transforms the original equation in a homogeneous first order
linear differential equation in the new variable  $z$,
$$
\frac{dz}{dt}=b_2(t)(x_1(t)-x_2(t))z \,,
$$
and therefore the general solution can be immediately found:
$$
z(t)=z(t=0)\exp\Big(\int_0^t b_2(s) (x_1(s)-x_2(s))\,ds\Big) \,.
$$

 Finally, if three particular solutions, $x_1,x_2,x_3$, are known,
 the general solution can be
 written, without making use of any quadrature, in the following way:
$$
\frac{x-x_1}{x-x_2}:\frac{x_3-x_1}{x_3-x_2}=k . 
$$
This is a
nonlinear superposition principle which has been studied  in
\cite{CMN} from a group theoretical perspective.

The simplest case of (\ref{ricceq}) being an autonomous equation ($b_0$, $b_1$ 
and $b_2$ constants), has been fully studied (see
e.g. \cite{CarRamdos} and references therein) and it is integrable by
quadratures. This is a consequence of the existence of a constant
(maybe complex) solution, which allows us to reduce the problem to an
inhomogeneous linear one. Also separable Riccati equations, of the form
$$
\frac{dx}{dt}=\varphi(t)(c_0+c_1 x+c_2 x^2)\, ,
$$
are integrable, because it is enough to introduce a new time variable $\tau $
such that $d\tau/dt=\varphi(t)$ and the problem reduces to an autonomous case.

It will be shown in Section 5  that there are other cases of Riccati equations
related with this type of separable or autonomous ones.

\section{Geometric Interpretation of Riccati equation}

 From the geometric viewpoint the Riccati equation
(\ref{ricceq})  can be considered as a differential equation
determining the integral curves of  the time-dependent vector
field
\begin{equation}
 \Gamma=(b_0(t)+b_1(t)x+b_2(t)x^2)\frac{\partial}{\partial x} \,.\label{vfRic}
\end{equation}
 The simplest
case is when all the coefficients $b_\alpha(t)$ are constant,
because then $\Gamma$, given by (\ref{vfRic}),
is a true vector field. Otherwise, $\Gamma$ is a vector field along
the projection map $\pi:\mathbb{R}\times \mathbb{R}\to \mathbb{R}$, 
given by $\pi(t,x)=x$
(see for example  \cite{Car96} where it is shown that these vector fields
 along $\pi$ also admit integral curves).

The important point is that $\Gamma$ is a linear combination with
time-dependent coefficients of the three  vector fields
\begin{equation}
L_0 =\frac{\partial}{\partial x} ,  \quad L_1 =x \frac{\partial}{\partial x}  , 
 \quad L_2 = x^2
\frac{\partial}{\partial x} ,  \label{sl2gen}
\end{equation}
which close on a 3-dimensional real
Lie  algebra, with defining relations
\begin{equation} \label{conmutL}
[L_0,L_1] = L_0 ,      \quad [L_0,L_2] = 2L_1 ,
  \quad [L_1,L_2] = L_2\, .
\end{equation}
Consequently this Lie algebra is isomorphic to
$T_ISL(2,\mathbb{R})$, considered as a Lie algebra in the natural way,
which is made up of traceless $2\times 2$ matrices. A particular
basis is given by
\begin{equation}
M_0=\begin{pmatrix}0&1\\ 0&0 \end{pmatrix}, \quad
M_1=\frac{1}{2}\begin{pmatrix} 1&0\\ 0&-1\end{pmatrix} ,\quad
M_2=\begin{pmatrix} 0&0 \\-1&0 \end{pmatrix} ,
\label{base_matrices}
\end{equation}
for which
$$
[M_0,M_1] = -M_0 ,  \quad [M_0,M_2] = -2M_1 ,     \quad
[M_1,M_2] = -M_2
$$
that are  commutation relations  analogous to (\ref{conmutL}), i.e.
the linear correspondence $L_\alpha\mapsto M_\alpha$ is
 an antihomomorphism of Lie algebras.

Note also that $L_0$ and $L_1$
generate a 2-dimensional Lie subalgebra isomorphic to the Lie algebra
 of the affine group of transformations in one dimension,
and the same holds for $L_1$ and $L_2$.
The one-parameter subgroups of local transformations of $\mathbb{R}$
generated  by  $L_0$, $L_1$ and  $L_2$ are
$$
x\mapsto x+\epsilon ,\quad x\mapsto e^\epsilon x ,\quad x
\mapsto\frac x{1-x\epsilon} \,.
$$

Note that $L_2$ is not a complete vector field on $\mathbb{R}$. However we can
do the one-point compactification of $\mathbb{R}$ and then
$L_0$, $L_1$ and  $L_2$ can be considered as the fundamental vector fields
corresponding to the action of $SL(2,\mathbb{R})$ on the completed real
line $\overline{\mathbb{R}}$, given by
\begin{gather*}
\Phi(A,x)={\frac{\alpha  x+\beta}{\gamma  x+\delta}},\quad
\mbox{if } x\neq-{\frac{\delta}{\gamma}}, \\
\Phi(A,\infty)={\alpha}/{\gamma} , \quad
\Phi(A,-{\delta}/{\gamma})=\infty, \quad
\mbox{when } A=\begin{pmatrix} \alpha & \beta \\
\gamma & \delta \end{pmatrix} \in SL(2,\mathbb{R}).
\end{gather*}

Let $\Phi:SL(2,\mathbb{R})\times\overline{\mathbb{R}}\to
\overline{\mathbb{R}}$ be the preceding  effective
left action
of $SL(2,\mathbb{R})$  on the one-point compactification of the real line. The
remarkable fact is the following: let $A(t)$ be
 the curve  in $SL(2,\mathbb{R})$ that is the integral curve of the 
 $t$-dependent vector field
$$
X(t,A)=-\sum_{\alpha=0}^2 b_\alpha(t)  X^{\tt R}_\alpha(A) ,
$$
which starts from the neutral element, i.e.  $A(0)=I$. Here
$\{M_\alpha\mid\alpha=0,1,2\}$ is a basis of the tangent
space $T_ISL(2,\mathbb{R})\simeq\mathfrak{sl}(2,\mathbb{R})$ and $ X^{\tt
  R}_\alpha$ denotes
the right-invariant vector field in $SL(2,\mathbb{R})$ such that $X^{\tt
  R}_\alpha(I)=M_\alpha$. This vector field satisfies that
   $X^{\tt R}_\alpha(A)=M_\alpha A$. In other words,
$A(t)$ satisfies
\begin{equation}\label{eLA}
\dot{A}(t)A^{-1}(t)=-\sum_{\alpha=0}^2b_\alpha(t)M_\alpha\equiv
{\rm a}(t) .
\end{equation}
Then, $x(t)=\Phi(A(t),x_0)$ is the  solution of the Riccati
equation (\ref{ricceq}) with initial condition $x(0)=x_0$. Also,
we remark that the r.h.s. of (\ref{eLA}) is a curve in
$T_ISL(2,\mathbb{R})$ that can be identified with a curve in the
Lie algebra $\mathfrak{sl}(2,\mathbb{R})$ of left-invariant vector
fields on $SL(2,\mathbb{R})$  through the usual isomorphism: we
relate a left-invariant vector field $X$ with the element $X(I)\in
T_ISL(2,\mathbb{R})$. From now on we will not distinguish explicitly
between elements in $T_ISL(2,\mathbb{R})$ and $\mathfrak{sl}(2,\mathbb{R})$.

In summary, the general solution of the  Riccati equation
(\ref{ricceq})  can be obtained through the curve in $SL(2,\mathbb{R})$
which starts from $I$ and it is solution of equation (\ref{eLA}).
Therefore we have transformed the problem of finding the general
solution of (\ref{ricceq}) to that of determining the curve
solution of (\ref{eLA}) starting from the neutral element.

\section{Transformation laws of Riccati equations}

In this section we describe an important property of Lie systems, in the
particular case of Riccati equation, which plays a very relevant r\^ole for
establishing, as indicated in \cite{CarRam},  integrability
criteria.  The group of curves in the group defining
the Lie system, here  $SL(2, \mathbb{R})$, acts on the set  of
Riccati equations.

More explicitly, each Riccati equation  (\ref{ricceq}) can be
considered as a curve in $\mathbb{R}^3$. The point now is that  we can
transform every function in $\overline{\mathbb{R}}$, $x(t)$, under an element of the
group of smooth $SL(2, \mathbb{R})$-valued curves $\mathop{\rm Map}(\mathbb{R},
SL(2,\mathbb{R}))$, which from now on will be denoted as $\mathcal{G} $, as
follows \cite{CarRam}:
\begin{gather}
\Theta(A,x(t))={\frac{\alpha(t) x(t)+\beta(t)}{\gamma(t) x(t)+\delta(t)}} ,\quad
\mbox{if } x(t)\neq-{\frac{\delta(t)}{\gamma(t)}} ,\label{acciondecurva}
\\
 \Theta(A,\infty)={\alpha(t)}/{\gamma(t)} ,\quad
\Theta(A,-{\delta(t)}/{\gamma(t)})=\infty ,\nonumber\\
\mbox{when } A=\begin{pmatrix} \alpha(t)& \beta(t) \\
\gamma(t) & \delta(t) \end{pmatrix} \in\mathcal{G} .\nonumber
\end{gather}
If we transform a curve $x(t)$ solution of (\ref{ricceq}) by means
of $x'(t)=\Theta({\bar A}(t),x(t))$, the function  $x'$ satisfies
a new Riccati equation with coefficients $b'_2,b'_1, b'_0$, given
by
\begin{equation} \label{trans}
\begin{aligned}
 b'_2&={\bar\delta}^2 b_2-\bar\delta\bar\gamma
b_1+{\bar\gamma}^2 b_0+\bar\gamma {\dot{\bar\delta}}-\bar\delta \dot{\bar\gamma} ,
 \\
b'_1&=-2 \bar\beta\bar\delta b_2+(\bar\alpha\bar\delta+\bar\beta\bar\gamma) b_1-2
\bar\alpha\bar\gamma b_0
       +\bar\delta \dot{\bar\alpha}-\bar\alpha \dot{\bar\delta}+\bar\beta \dot{\bar\gamma}
-\bar\gamma \dot{\bar\beta} ,   \\
b'_0&= {\bar\beta}^2 b_2-\bar\alpha\bar\beta b_1+{\bar\alpha}^2
b_0+\bar\alpha\dot{\bar\beta}-\bar\beta\dot{\bar\alpha}  .
\end{aligned}
\end{equation}

In fact  this expression defines an affine action (see e.g.  \cite{LM87}
for the definition of this concept) of the group
 $\mathcal{G} $ on the set of Riccati equations. More details can be found
in \cite{CarRam}.
This means that in order  to transform the coefficients of a
general Riccati equation by means of first a transformation given
by the curve $A_1(t)$ and then another defined by the curve
$A_2(t)$, it suffices to do the transformation by the product
element $A_2 A_1$ of $\mathcal{G} $.

The result of the action of $\mathcal{G}$ can also be studied from
the point of view of the equations in $SL(2,\mathbb{R})$.
First, $\mathcal{G}$  acts on the left on the
set of curves in $SL(2, \mathbb{R})$ by left translations, i.e. a curve $\bar A(t)$
transform the curve $A(t)$  into a new one  $A'(t)=\bar A(t) A(t)$, and if
 $A(t)$ is a solution of (\ref{eLA})
then the new curve satisfies a new equation like (\ref{eLA}) but
with a different right hand side, and therefore it corresponds
to a new equation in $SL(2,\mathbb{R})$ associated with a new Riccati equation.
In this way $\mathcal{G}$
 acts on the set of curves in $\mathfrak{sl}(2,\mathbb{R})$, i.e. on the set of
 Riccati equations. It can be shown that
 the relation between both
curves in $T_{I}SL(2,\mathbb{R})$ is given by:
\begin{equation}
{\rm a}'(t)=\bar A(t){\rm a}(t)\bar A^{-1}(t)+\dot{\bar{A}}(t)\bar
A^{-1}(t) =-\sum_{\alpha=0}^2 b'_\alpha(t)M_\alpha .
\label{newricc}
\end{equation}
Therefore,  $A(t)$ is a solution of (\ref{eLA})
 if and only if  $A'(t)=\bar A(t) A(t)$ is  a solution for the equation
corresponding to the curve ${\rm a}'(t)$ given by (\ref{newricc}).
If we restrict ourselves (\ref{eLA}) to curves starting from the
neutral element, $\bar A(t)$ should also start from $I$. The
transformed  Riccati equation, obtained through (\ref{trans}) is
the one related with equation (\ref{newricc}) in the group
$SL(2,\mathbb{R})$.

We have shown that it is possible to associate Riccati equations
with  equations in the Lie group $SL(2,\mathbb{R})$ and to define
a group of transformations on the set of Riccati equations.
Roughly speaking, we have transformed the initial problem of
Riccati equations on $\overline{\mathbb{R}}$ into a problem in the group
$SL(2,\mathbb{R})$ and we have explained a way to relate both problems.

\section{Interpretation of integrability conditions}

In this section we analyze and reproduce the results of \cite{
Ra61, AS64, Ra62,RU68,Ko06,RDM05} from our geometric viewpoint.
Our approach  provides us with additional information about  why
the methods used in these papers work.

In summary, in most of these papers one starts with a Riccati
differential equation
\begin{equation}\label{initRiccati}
 \frac {dy}{dt}=b_0(t)+b_1(t)y+b_2(t)y^2
\end{equation}
such that  under  a time-dependent change of variables $y'\equiv y'(y,t)$ 
the initial equation is transformed into
\begin{equation}\label{fe}
 \frac {dy'}{dt}=\varphi(t)(c_0+c_1y'+c_2 y'^2)
\end{equation}
with constants $c_0, c_1$ and $c_2$. This new equation is integrable as
mentioned before. Then
the solution of the initial differential equation is obtained from
the solution of (\ref{fe}) in terms of the initial variable.

The key point in this method is that of finding out an appropriate
 time-dependent change of
coordinates in order to transform the initial Riccati equation
into a  new one that can be integrated. Then, it is interesting to
know the possible forms of these time-dependent changes of
variables and  their mathematical meaning.

Now, we will explain how to use our geometrical formalism for
finding the general solution of the given  Riccati equations and
in particular for explaining the results of different
 papers about this topic in the
literature.

The process of finding out the general  solution for a Riccati
 equation  can be seen   as similar to a reduction process as it appears in
 \cite{CMN, CarRamGra}.
In our geometrical formalism, given an initial Riccati equation,
there  is a related equation in the group $SL(2,\mathbb{R})$ characterized
by a curve ${\rm a}(t)$ in the Lie algebra of $SL(2,\mathbb{R})$.
If this equation is not integrable, we can try to transform it
into a new Riccati equation related with a Lie  subalgebra
$\mathfrak{h}\subset {\mathfrak{sl}}(2,\mathbb{R})$ which may be  either  a
1-dimensional one given by $\langle c_0 M_0+c_1M_1+c_2M_2\rangle $, with
$c_0,c_1,c_2$ fixed real numbers, or a 2-dimensional one,
isomorphic to the Lie algebra of the affine group on the real
line. This transformation  process is carried out  by means of the
action on the given  equation in the group of an appropriate
curve $\bar A(t)$ in such group. More specifically, this $\bar
A(t)$ must be such that:
\begin{equation}
{\rm a}'(t)=\bar A(t){\rm a}(t)\bar A^{-1}(t)+\dot {\bar A}(t)\bar
A^{-1}(t) =-\varphi(t)\sum_{\alpha=0}^2c_\alpha M_\alpha\in T_IH
\label{eq_ap}
\end{equation}
This is a Lie system in the  connected Lie subgroup $H\subset
SL(2,\mathbb{R})$ with Lie algebra $\mathfrak{h}\subset \mathfrak{g}$. 
Once such a
curve $\bar A(t)$ has been found, and also the solution of the
corresponding Riccati equation to ${\rm a}'(t)$ given by
(\ref{eq_ap}), we can obtain the solution of the original equation
in the group $SL(2,\mathbb{R})$ and $\overline{\mathbb{R}}$ by
inverting the transformation carried out by $\bar A(t)$. In other
words, if $A'(t)$ and $x'(t)$ are the solutions of the transformed
equations, then $A(t)=\bar A^{-1}(t) A'(t)$ and $x(t)=\Theta(\bar
A^{-1}(t),x'(t))=\Theta(\bar A^{-1}(t)A'(t),x_0)$ are the
solutions of the initial equations in the group $SL(2,\mathbb{R})$ and in
$\overline{\mathbb{R}}$, respectively.

The important point here is that the described
action of the group $\mathcal{G}$ of curves in
$SL(2,\mathbb{R})$   on the set of Lie systems in the group $SL(2,\mathbb{R})$
 (see
\cite{CarRam05b} and \cite{Car96}) can be used to construct transformations in
the set of
Riccati equations that are general enough to reproduce the
 transformations that have been used  in the literature.

We next reproduce   some results of the papers
\cite{ Ra61,AS64,Ra62,RU68,Ko06,Zh99} from  our geometrical approach.
Consider first  the example studied in \cite{Ra62} where  the following
Riccati equation
\begin{equation}\label{inRicc}
\frac{dy}{dt}=b_0(t)+b_1(t)y+b_2(t)y^2,\quad b_2(t)>0
\end{equation}
is analyzed in an interval such that
$W=b_2^2(t)b_0(t)+\dot b_1(t)b_2(t)-b_1(t)\dot b_2(t)>0$, where
\begin{equation}\label{eqW}
\frac{b_2(t)W'-(3\dot b_2(t)-2
b_1(t)b_2(t))W}{2b_2(t)^{1/2}W^{3/2}}\equiv K ,
\end{equation}
$K$ being a constant.
Defining $v$ by putting $W=b_2(t)^3v^2$ and using the so-called
Rao's transformation:
$$
y=v(t)y'-b_1(t)/b_2(t) ,
$$
 the Riccati equation (\ref{inRicc}) is transformed into
$$
\frac{dy'}{dt}=\varphi(t)(c_0+c_1 y'+c_2 y'^2) ,
$$
where $c_0,c_1$ and $c_2$ are real numbers  satisfying
\begin{gather*}
\varphi(t) c_0=\sqrt{\frac{W(t)}{b_2(t)}},\\
\varphi(t) c_1=-K\sqrt{\frac{W(t)}{b_2(t)}},\\
\varphi(t) c_2=\sqrt{\frac{W(t)}{b_2(t)}}.
\end{gather*}
Such new Riccati equation is separable and therefore
it  is integrable.

In the framework of Lie systems theory we start with the associated
equation in $SL(2,\mathbb{R})$ given by
$$
\dot A(t)A^{-1}(t)=-\sum_{\alpha=0}^2 b_\alpha(t) M_\alpha
={\rm a}(t) ,
$$
which under the left action of  a curve
$\bar A(t)\in \mathop{\rm Map}(\mathbb{R},SL(2,\mathbb{R}))$
 becomes a similar new equation but where the
curve ${\rm a}(t)$ is replaced by  a  new
 one ${\rm a}'(t)$ according to formulas (\ref{trans}) or (\ref{newricc}).
In particular,  with the  choice
$$
\bar\alpha(t)=\frac{1}{\sqrt{v(t)}}, \quad
\bar\beta(t)=\frac{b_1(t)}{b_2(t)\sqrt{v(t)}},\quad
\bar\gamma(t)=0,\quad
\bar\delta(t)=\sqrt{v(t)} ,
$$
we define a transformation that under the conditions
 imposed in the article gives the following result:
$$
\frac{dy'}{dt}=\Big( \frac{W}{b_2(t)} \Big)^{1/2}(1-Ky'+y'^2) .
$$
This equation corresponds to the final Riccati equation found in
\cite{Ra62}. A similar transformation is that of Theorem~2 of \cite{RU68}.
The initial
 Riccati equation associated with an equation in the group $SL(2,\mathbb{R})$
 characterized by a curve ${\rm a}(t)$ in $\mathfrak{sl}(2,\mathbb{R})$,  
 becomes  a new Lie system
 under
 the proposed transformation
  but now characterized by a curve in the 1-dimensional Lie subalgebra
  $\mathfrak{h}=\langle M_0-KM_1+M_2\rangle$.
The equation in the subgroup determined by $\mathfrak{h}$ is integrable, 
and once its solution is known
we recover the solution of the initial problem in $SL(2,\mathbb{R})$ as
$A(t)=\bar A^{-1}(t)A'(t)$. Finally, the solution for the initial Riccati
equation with initial condition $y(0)=y_0$ is
given by  $y(t)=\Theta(\bar{A}^{-1}(t) A' (t),y_0)$.

Another case also studied in \cite{Ra62} is when $W=0$. In this case,
 we can perform the transformation defined by
\begin{align*}
\bar\alpha(t)&=\exp\Big(\frac 12 \int_0^tb_1(t')dt'\Big), \\
\bar\beta(t)&=\exp\Big(\frac 12 \int_0^tb_1(t')dt'\Big)\frac{b_1(t)}{b_2(t)},\\
\bar\gamma(t)&=0 \\
\quad \bar\delta(t)&=\exp\Big(-\frac 12\int_0^tb_1(t')dt'\Big),
\end{align*}
and then, the  new  equation is
$$
\frac{dy'}{dt}=b_2(t)  \bar\alpha^{-2}(t) y'^2 .
$$
In this way we transform the initial Riccati equation, with $W=0$,
into a new one characterized by a curve in the one dimensional
Lie subalgebra $\mathfrak{h}=\langle M_2\rangle \subset \mathfrak{sl}(2,\mathbb{R})$, and we can solve the
problem
as before. This last equation is also obtained in \cite{Ra62}.

Another case of integrable Riccati equation (\ref{ricceq}) is considered
in \cite{RU68}, where the coefficient functions $b_0(t),b_1(t),b_2(t)$
are assumed to satisfy the relations
$$
\frac{dv}{dt}=-kb_0(t)+b_1(t)v,\quad
b_2(t)=\frac{b_0(t)}{cv^2(t)},
$$
where $v(t)$ is a new function and $c$ and $k$ are two real constants.
We can transform (\ref{ricceq}) by means of  the curve $\bar A(t)$
of coefficients
$$
\bar\alpha(t)=\frac{1}{\sqrt{v(t)}},\quad
\bar\beta(t)=0,\quad\bar\gamma(t)=0,\quad
\bar\delta(t)=\sqrt{v(t)},
$$
into
$$
\frac{dy'}{dt}=\frac{b_0(t)}{v(t)}\Big(\frac{y'^2}{c}+ky'+1\Big).
$$

Another example, given in  \cite{AS64}, is a Riccati equation
of form (\ref{ricceq}) whose coefficient functions $b_0(t),b_1(t),b_2(t)$
satisfy
\begin{equation}
\frac{b_1(t)+\frac{1}{2}\left(\frac{\dot b_2(t)}{b_2(t)}-\frac{\dot
b_0(t)}{b_0(t)}\right)}{\sqrt{b_0(t)b_2(t)}}=C , \label{eqC}
\end{equation}
where $C$ is a constant. In such a case it can be transformed into
the new one:
$$
\frac{dy'}{dt}=\sqrt{b_0(t)b_2(t)}(1+Cy'+y'^2)
$$
by means of the transformation given by the curve $\bar A(t)$ in 
$SL(2,\mathbb{R})$ with
$$
\bar\alpha(t)=\Big(\frac{b_2(t)}{b_0(t)}\Big)^{1/4},\quad
\bar\beta(t)=0,\quad\bar\gamma(t)=0,\quad
\bar\delta(t)=\Big(\frac{b_0(t)}{b_2(t)}\Big)^{1/4}.
$$

Another example that can  be analyzed from  our viewpoint is
the following Riccati equation, \cite{Ko06},
$$
\frac {dy}{dt}=-\frac {c_1}{F(t)}y^2+\Big(c_2+\frac
{F'(t)}{F(t)}\Big)y+F(t) .
$$
In this case, if $F(t)>0$ we can perform the transformation defined by
$$
\bar\alpha(t)=\sqrt{1/F(t)}, \quad \bar\beta(t)=0, \quad
\bar\gamma(t)=0,\quad
\bar\delta(t)=\sqrt{F(t)} ,
$$
and the following new Riccati equation is obtained:
$$
\frac {dy'}{dt}=-c_1y'^2+c_2 y'+1,
$$
which can  easily be integrated. This is a particular case of the example in
 \cite{RU68} explained above, with $v(t)=F(t)$ , $k=c_2$ and $c=-1/c_1$.
Once again, we have performed a transformation  from the initial
problem in $SL(2,\mathbb{R})$ characterized by a curve in the Lie algebra
$\mathfrak{sl}(2,\mathbb{R})$ into a new equation in the 1-dimensional Lie
subgroup characterized by a curve ${\rm a}'(t)$ in the
1-dimensional subalgebra $\mathfrak{h}=\langle M_0+c_2M_1-c_1M_2\rangle $. 
The solution
$A'(t)$ with initial condition $A'(0)=I$ lies in the corresponding
subgroup and  the solution to the initial Riccati equation can be
obtained from it as indicated before.

Other transformations that  can be used and the corresponding integrable
Riccati equations are:
\begin{gather*}
\bar A(t)=\begin{pmatrix}
\sqrt{-\frac{c_1}{F(t)}}&0\\
0&\sqrt{-\frac{F(t)}{c_1}}
\end{pmatrix}
\Longrightarrow
\frac{dy'}{dt}=y'^2+c_2 y'-c_1,\\
\bar A(t)=\begin{pmatrix} \Big(\frac{-c_1}{F^2(t)}\Big)^{1/4}&0\\0&
\Big(-\frac{F^2(t)}{c_1}\Big)^{1/4}
\end{pmatrix}
\Longrightarrow
\frac{dy'}{dt}=\sqrt{-c_1}y'^2+c_2 y'+\sqrt{-c_1}
\end{gather*}
The first case is a particular case of the cited example in \cite{RU68}
with $v(t)=-F(t)/c_1$, $k=-c_2/c_1$ and $c=-c_1$. The second case
is a particular case of the example in \cite{AS64} developed above,
where the constant $C$ in (\ref{eqC}) takes the value $c_2/\sqrt{-c_1}$.
In both cases  the time-dependent changes of coordinates can
be described through our set of transformations.

We next study  from our perspective an  example which can be found
in \cite{Ra61}. According to its results the Riccati equation can
be  integrated directly if the time-dependent
 coefficients satisfy the relation
$$
\frac{d}{dt} \log\frac{-b_0(t)}{b_2(t)}= 2b_1(t).\\
$$
or, equivalently,
$$
\log\frac{-b_0(t)}{b_2(t)}= 2\int^t_0 {b}_1(t')dt'+\log a ,
$$
where $a>0$ is a constant.
In this case, the way to proceed is just by means of the action
(\ref{acciondecurva}) of the
curve in $SL(2,\mathbb{R})$ given by
\begin{gather*}
\bar\alpha(t)=\exp\Big(-\frac 12\int^t_0 {b}_1(t')dt'\Big),\quad
\bar\beta(t)=0,\\\
\bar\gamma(t)=0, \quad
\bar\delta(t)=\exp\Big(\frac 12\int^t_0 {b}_1(t')dt'\Big).
\end{gather*}
The final result is
$$
\frac{dy'}{dt}=(y'^2-a)\bar\alpha^{-2}(t)b_2(t),
$$
which can be integrated and thus, by inverting the time-dependent change
of variables, we obtain the solution for the initial Riccati equation.

The aim of the transformations on Riccati equations considered so
far was to obtain separable Riccati equations. This is not the
only way to integrate such differential equations, but we can also
transform the given Riccati equation into another one associated,
as Lie system, with a bidimensional real Lie algebra, which is
solvable, and therefore   the corresponding Lie system can
integrated in this way \cite{CarRamGra}.

A particular instance of this which can be found in \cite{RDM05}
is
$$
\frac{dy}{dt}=P(t)+Q(t)y+k(Q(t)-kP(t))y^2 ,
$$
where $k$ is a non-vanishing constant. If we consider the
transformation (\ref{acciondecurva}) with $\bar\alpha(t)=0,
\bar\beta(t)=-1/k,\bar\gamma(t)=k,\bar\delta(t)=1$, it gives rise
to the time independent change of variables
\begin{equation}
y'=-\frac{1}{k^2y+k} ,\label{solconst}
\end{equation}
under which we obtain an inhomogeneous
linear  equation
\begin{equation}
\frac{dy'}{dt}=\frac{Q(t)}{k}-P(t)+(Q(t)-2kP(t))y'  .
\label{equation_lin_hom}
\end{equation}
Note that this transformation (\ref{solconst}) corresponds to the fact
that $y=-1/k$ is a solution
of the original Riccati equation.

In this equation the dynamical vector field
$$
X=\Big(\frac{Q(t)}{k}-P(t)\Big)\frac{\partial}{\partial y'}+(Q(t)
- 2kP(t))y'\frac{\partial}{\partial y'}
$$
can be written in terms of two vector fields in $\mathbb{R}$
$$
X_1=\frac{\partial}{\partial y'},\quad X_2=y'\frac{\partial}{\partial y'}  ,
$$
because
 $$
X=\Big(\frac{Q(t)}{k}-P(t)\Big)X_1+(Q(t)-2kP(t))X_2 .
$$
 These vector fields are such that $[X_1,X_2]=X_1$
 and thus they close on a bidimensional non-Abelian real
Lie algebra, which, as it is well-known,  is solvable and
(\ref{equation_lin_hom}) can be integrated by quadratures.

Finally, all the examples given in \cite{Zh99} can be dealt with  our
geometric formalism.
 Next, we shall specify some of the cases of such paper. The first one is a
 Riccati equation (\ref{ricceq}) such that the time-dependent coefficients
 satisfy that for a certain
function $D(t)$ and constants $a,b,c$:
\begin{gather*} 
b_2(t) b_0(t)=a c D^2(t) ,\\
\frac{\dot b_2(t)}{b_2(t)}+b_1(t)=\frac{\dot D(t)}{D(t)}+b D(t) .
\end{gather*}
It can be noted that if $a,b,c$ and $D(t)$ satisfy the above
integrability conditions
 in an interval of $t$, then $-a,b,-c$ and $D(t)$ satisfy the
 same conditions in the same interval. Then, we can always choose $a$
in such a way that ${a D(t)}/{b_2(t)}>0$ in that interval. Thus,
the transformation given by $\bar A(t)$ with coefficients:
$$
\bar\alpha(t) =\sqrt{\frac{b_2(t)}{a D(t)}},\quad\bar\beta(t)=0,\quad
\bar\gamma(t)=0,\quad
\bar\delta(t)=\sqrt{\frac{a D(t)}{b_2(t)}} ,
$$
allows us to transform the initial Riccati equation  into
$$
\frac{dy'}{dt}=D(t)(c+b y'+a y'^2) .
$$
We have transformed the associated equation in the Lie group
$SL(2,\mathbb{R})$ to the initial Riccati equation into a new one in a
1-dimensional Lie subgroup of $SL(2,\mathbb{R})$ of Lie algebra given by
$\mathfrak{h}=\langle c M_0+b M_1+a M_2\rangle$.

Other integrability conditions proposed in \cite{Zh99} are:
\begin{gather*} 
b_2(t) \Big(-\frac{dE}{dt}+b_2(t)E^2(t)+b_1(t)E(t)+b_0(t)\Big)
=a c D^2(t) ,\\
\frac{\dot b_2(t)}{b_2(t)}+b_1(t)+2E(t) b_2(t)=\frac{\dot
D(t)}{D(t)}+b D(t) ,
\end{gather*}
where $a,b,c$ are constants and $D(t)$, $E(t)$ are functions.
As before, we can always choose $a$ in such a way that
 ${a D(t)}/{b_2(t)}>0$. Then, if we consider the transformation given 
 by $\bar A(t)$, with coefficients:
\begin{equation*}
\bar\alpha(t)=\sqrt{\frac{b_2(t)}{a D(t)}},\quad
\bar\beta(t)=-\sqrt{\frac{b_2(t)}{a D(t)}}E(t),\quad
\bar\gamma(t)=0,\quad
\bar\delta(t)=\sqrt{\frac{a D(t)}{b_2(t)}} ,
\end{equation*}
we see that the initial Riccati equation transforms into
\begin{equation*} 
\frac{dy'}{dt}=D(t)(c+b y'+a y'^2) ,
\end{equation*}
which  is integrable, and then  we can obtain the solution to
the initial Riccati equation.
Other results of \cite{Zh99} are summarized in  Table \ref{table1}.

\begin{table}[ht]
\caption{Some integrability conditions in \cite{Zh99}} \label{table1}
\begin{center}
 \begin{tabular}{|c|c|}
\hline
Integrability condition & Transformation \\
\hline 
{\tiny $\begin{gathered}
b_2(t)b_0(t)=a c D^2(t),\\
\frac{\dot b_0(t)}{b_0(t)}-b_1(t)=\frac{\dot D(t)}{D(t)}-b D(t)
\end{gathered}$}
 &
{\tiny $\begin{pmatrix}
\sqrt{\frac{c D(t)}{b_0(t)}}&0\\
0&\sqrt{\frac{b_0(t)}{c D(t)}}
\end{pmatrix}$}
   \\ \hline
{\tiny $\begin{gathered}
b_2(t)L[E(t)]=a c D^2(t),\\
\frac{\dot L[E(t)]}{L[E(t)]}-b_1(t)-2 E(t) b_2(t)
=\frac{\dot D(t)}{D(t)}+b D(t)
\end{gathered}$}
 & {\tiny $\begin{pmatrix}
0&\sqrt{\frac{L[E(t)]}{a D(t)}}\\
-\sqrt{\frac{a D(t)}{L[E(t)]}}&E(t)\sqrt{\frac{a D(t)}{L[E(t)]}}
\end{pmatrix}$}
   \\ \hline
{\tiny $\begin{gathered}
b_2(t)L[E(t)]=a c D^2(t),\\
\frac{\dot b_2(t)}{b_2(t)}+b_1(t)+2 E(t) b_2(t)=\frac{\dot D(t)}{D(t)}-b D(t)
\end{gathered}$}
&
{\tiny $\begin{pmatrix}
0&\sqrt{\frac{c D(t)}{b_2(t)}}\\
-\sqrt{\frac{b_2(t)}{c D(t)}}&E(t)\sqrt{\frac{b_2(t)}{c D(t)}}
\end{pmatrix}$}
\\ \hline
{\tiny $\begin{gathered}
b_2(t)L[E(t)]=a c D^2(t),\\
\frac{\dot L[E(t)]}{L[E(t)]}-b_1(t)-2 E(t) b_2(t)
=\frac{\dot D(t)}{D(t)}-b D(t)
\end{gathered}$}
 & {\tiny $\begin{pmatrix}
\sqrt{\frac{c D(t)}{L[E(t)]}}&-E(t)\sqrt{\frac{c D(t)}{L[E(t)]}}\\
0&\sqrt{\frac{L[E(t)]}{c D(t)}}
\end{pmatrix}$}
   \\ \hline
{\tiny $\begin{gathered}
\Big(\frac{B(t)}{A(t)}\Big)^2L[E(t)]L2[E(t)]=a c D^2(t),\\
\frac{\dot L[E(t)]}{L[E(t)]}-\frac{2 B(t)}{A(t)}L[E(t)]
-b_1(t)-2 E(t) b_2(t)\\
=\frac{\dot D(t)}{D(t)}+b D(t)
\end{gathered}$}
 & {\tiny $\begin{pmatrix}
-\sqrt{\frac{L[E(t)]}{a D(t)}}\frac{B(t)}{A(t)}
&\sqrt{\frac{L[E(t)]}{a D(t)}}\Big(1+E(t) \frac{B(t)}{A(t)}\Big)\\
-\sqrt{\frac{a D(t)}{L[E(t)]}}&\sqrt{\frac{a D(t)}{L[E(t)]}} E(t)
\end{pmatrix}$}
\\ \hline
{\tiny $\begin{gathered}
\Big(\frac{B(t)}{A(t)}\Big)^2L[E(t)]L2[E(t)]=a c D^2(t),\\
\frac{\dot L2[E(t)]}{L2[E(t)]}
-\frac{2\frac{d}{dt}{\big(A(t)/B(t)\big)}}{A(t)/B(t)}
+\frac{2 B(t)}{A(t)}L[E(t)]\\
+b_1(t)+2 E(t) b_2(t)=\frac{\dot D(t)}{D(t)}-b D(t)
\end{gathered}$}
 &
{\tiny $\begin{pmatrix}
-\sqrt{\frac{c D(t)}{L2[E(t)]}}&\sqrt{\frac{c D(t)}{L2[E(t)]}}
\Big(1+\frac{B(t)}{A(t)} E(t)\Big)\frac{A(t)}{B(t)}\\
- \frac{B(t)}{A(t)}\sqrt{\frac{L2[E(t)]}{c D(t)}}&
\frac{B(t)}{A(t)}E(t)\sqrt{\frac{L2[E(t)]}{c D(t)}}
\end{pmatrix}$}
\\
\hline
\end{tabular}
\end{center}
\end{table}

In this table we have used:
\begin{gather*} 
L[E(t)]=-\frac{dE}{dt}+b_2(t)E^2(t)+b_1(t)E(t)+b_0(t), \\
L2[E(t)]=L\left[\frac{A(t)}{B(t)}+E(t)\right] .
\end{gather*}
As it happens in the examples above of \cite{Zh99}, the integrability
conditions presented in the table allow to change
$a\to -a, c\to -c$ leaving them invariant.
This symmetry is used implicitly in order to get the square
roots to be real.

It is to be remarked that some of the time-dependent transformations used
in \cite{Zh99} are homographies of the type
$$
y'=\frac{\alpha(t) y+\beta(t)}{\gamma(t) y+\delta(t)} ,
$$
for which the coefficients satisfy $\alpha(t)\delta(t)-\beta(t)\gamma(t)<0$
and thus they cannot be treated directly by the method presented here.
For example, a transformation like
$$
y'=\frac{A(t)b_2(t)}{a D(t)}y,\quad\text{with }
\frac{A(t)b_2(t)}{a D(t)}<0
$$
belong to this type. However, we can consider then a transformation 
$y\to y''$ of the form
$y''= -y$  and after the new transformation $y''\to y'$ given by
$$
y'=-\frac{A(t)b_2(t)}{a D(t)}y'',\quad\text{with }
-\frac{A(t)b_2(t)}{a D(t)}>0 ,
$$
 can be written as a homography with
$\alpha(t)\delta(t)-\beta(t)\gamma(t)=1$,
namely
$$
y'=\frac{\sqrt{-\frac{A(t)b_2(t)}{a D(t)}}y''}{\sqrt{-\frac{a
D(t)}{A(t)b_2(t)}}} =-\frac{A(t)b_2(t)}{a D(t)}y'' .
$$
In summary, we may use the set of time-dependent changes of
variables generated by curves
in the group $SL(2, \mathbb{R})$ in order to transform a given
Riccati equation into another one which is a Lie system with an
associated solvable Lie algebra, i.e. either 1-dimensional or
non-Abelian 2-dimensional one. In this way, the transformed Riccati equation
can be integrated by quadratures and thus,
 by the time-dependent change of variables, we can obtain the solution to the
initial Riccati equation.

\subsection*{Conclusions and outlook}
In summary it has been shown in this paper that previous works about the 
integrability of the Riccati
equation can be explained from the unifying viewpoint of Lie systems, and so
 appropriate transformations of $SL(2,\mathbb{R})$ can be used to transform
the
Riccati equations considered in the mentioned  papers
 into other ones that can be easily integrated. These transformations are 
made in such a way that the initial
Riccati equation, as a Lie system related with an equation in 
$SL(2,\mathbb{R})$, is
transformed into another Riccati equation related with an equation in a Lie
subgroup of $SL(2,\mathbb{R})$ with solvable algebra. The system is transformed in this way
into a simpler one and the integrability conditions arise  in this framework
as sufficient conditions for the existence of the convenient curve in
$SL(2,\mathbb{R})$ to be used in order to carry out this transformation process.

Another interesting problem to be studied in the future is
 to invert this process, i.e. we start  from a Riccati equation related with
 a solvable group as a Lie system and  perform a certain kind of
 transformations
by means of the action of curves of $SL(2,\mathbb{R})$ in order  to obtain
 Riccati equations integrable by quadratures. The integrability conditions 
appear  then
as properties of the set of Riccati equations obtained in this process.

\subsection*{Acknowledgements}
 Partial financial support by research projects MTM2006-10531 and E24/1 (DGA)
 is acknowledged. J.d.L. also acknowledges
 a F.P.U. grant from  Ministerio de Educaci\'on y Ciencia.

\end{document}